\documentclass[a4paper]{article}

\usepackage{INTERSPEECH2022}
\usepackage{cite}

\newcommand{\B}{\mathbf}
\newcommand{\BB}{\mathbb}

\usepackage{multirow}
\usepackage{array}
\usepackage{tabularx,booktabs}
\usepackage[textwidth=0.5in]{todonotes}
\usepackage{xcolor}

\newcolumntype{Y}{>{\centering\arraybackslash}X}

\title{Ultra-Low-Bitrate Speech Coding with Pretrained Transformers}
\name{Ali Siahkoohi\textsuperscript{1}, Michael
Chinen\textsuperscript{2}, Tom Denton\textsuperscript{2}, W. Bastiaan
Kleijn\textsuperscript{2,3}, and Jan Skoglund\textsuperscript{2} \thanks{This work was done during a
research internship at Google.}}
\address{\textsuperscript{1}School of Computational Science and
  Engineering, Georgia Institute of Technology\\
  \textsuperscript{2}Chrome Media, Google\\
  \textsuperscript{3}School of Engineering and Computer Science,
Victoria University of Wellington}
\email{alisk@gatech.edu, mchinen@google.com, tomdenton@google.com,
bastiaan.kleijn@ecs.vuw.ac.nz, jks@google.com}

\begin{document}

\maketitle
\begin{abstract}
Speech coding facilitates the transmission of speech over
low-bandwidth networks with minimal distortion. Neural-network based
speech codecs have recently demonstrated significant improvements in
quality over traditional approaches. While this new generation of
codecs is capable of synthesizing high-fidelity speech, their use of
recurrent or convolutional layers often restricts their effective
receptive fields, which prevents them from compressing speech
efficiently. We propose to further reduce the bitrate of neural speech
codecs through the use of pretrained Transformers, capable of
exploiting long-range dependencies in the input signal due to their
inductive bias. As such, we use a pretrained Transformer in tandem
with a convolutional encoder, which is trained end-to-end with a
quantizer and a generative adversarial net decoder. Our numerical
experiments show that supplementing the convolutional encoder of a
neural speech codec with Transformer speech embeddings yields a speech
codec with a bitrate of $600\,\mathrm{bps}$ that outperforms the
original neural speech codec in synthesized speech quality when
trained at the same bitrate. Subjective human evaluations
suggest that the quality of the resulting codec is
comparable or better than that of conventional codecs operating at
three to four times the rate.
\end{abstract}

\section{Introduction}
\label{introduction}

Speech compression aims to reduce the bitrate required to represent a
speech signal. In classical coding
methods~\cite{rcelp1994, bielefeld1996developing, mccree19962,
melpe2002, valin2012definition, dietz2015overview, valin2016speex}, all
processing was based on knowledge of human experts only. Recent
advances in speech coding follow progress in speech synthesis
\cite{oord2016wavenet, kumar2019melgan, kong2020hifi} by replacing the decoder~\cite{kleijn2018wavenet,
valin2019lpcnet, Klejsa2019} as well as the quantizer~\cite{garbacea2019low} with a
machine-learning (ML) based model that significantly
improves the coding quality. More recently, end-to-end coding schemes
have been developed~\cite{lim2020robust,zeghidour2021soundstream} that employ an autoencoder structure with quantization in the bottleneck (latent space).
With SoundStream~\cite{zeghidour2021soundstream}, this autoencoding structure, in the form of
a VQ-VAE~\cite{oord2018representation}, was further combined with the
learned distortion measures from generative
adversarial networks (GANs)~\cite{goodfellow2014generative}.
While~\cite{zeghidour2021soundstream} represents the current
state-of-the-art above $3\,\mathrm{kbps}$, its relative effectiveness
deteriorates at lower rates. Indications exist~\cite{jafarlou2019analyzing} that lengthening the effective receptive field
of the encoder may improve performance at very low rates. This motivates
us to study the combination of the approach
of~\cite{zeghidour2021soundstream} with an encoder that can exploit
long-term dependencies in the input speech signal.

Our work focuses on reducing the bandwidth of neural speech codecs by
incorporating speech embeddings that are obtained from a pretrained
Transformer model~\cite{baevski2020wav2vec, gulati2020conformer,
zhang2020pushing}. The Transformer is pretrained in a self-supervised learning
framework~\cite{baevski2020wav2vec, zhang2020pushing} that involves performing a contrastive
task~\cite{oord2018representation} over large quantities of raw speech. Owing to their multi-head attention
layers~\cite{vaswani2017attention}, Transformers have the inductive bias to exploit
long-distance dependencies in the input speech, which enables them to learn speech embeddings that
result in state-of-the-art performance on speech-related downstream
tasks~\cite{polyak2021speech, gulati2020conformer, baevski2020wav2vec,
zhang2020pushing, chung2021w2v}.

\begin{figure}
  \centering
  \includegraphics[width=1.0\hsize]{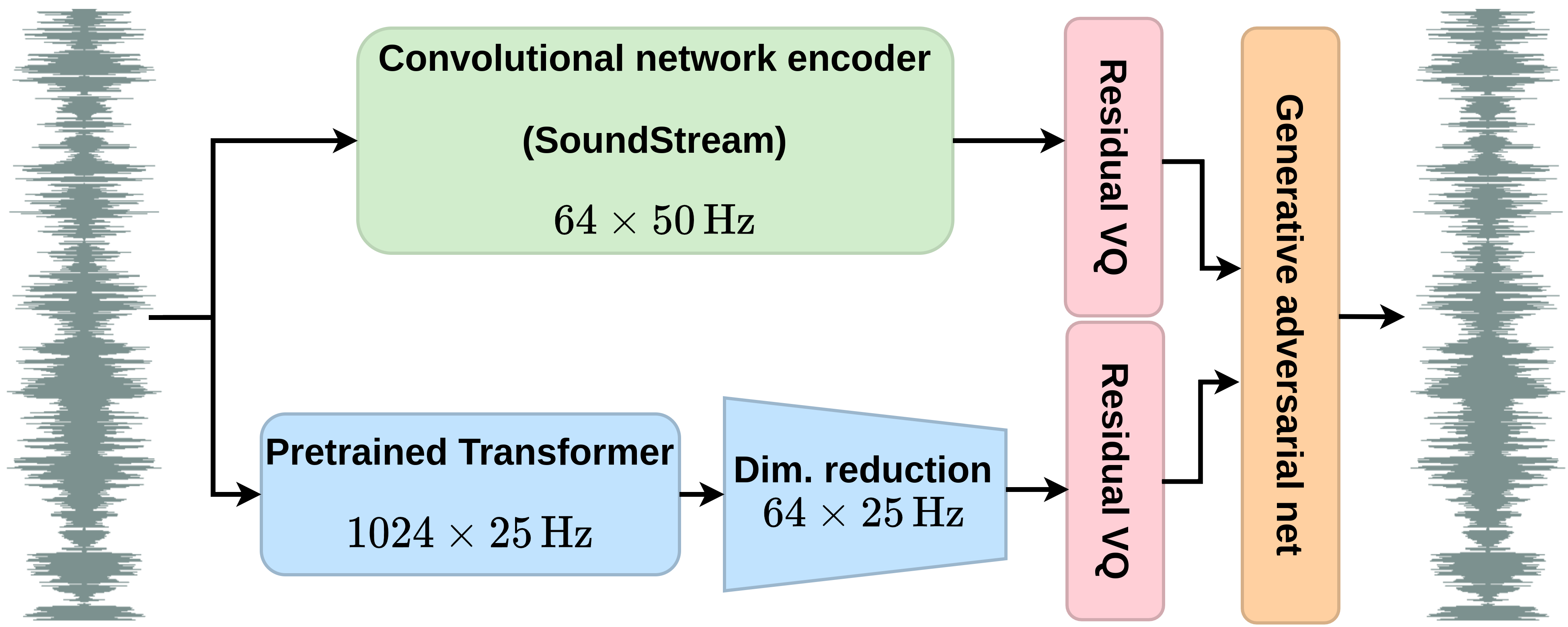}
  \caption{The proposed speech codec. Speech is encoded via Transformer
  embeddings and SoundStream-based CNN encoder features. After
  quantization with residual VQs, a GAN-based decoder reconstructs the
  input speech.}
  \vspace*{-0.5cm}
  \label{proposed_methpd}
\end{figure}

To design the encoder in our ultra-low-bitrate codec, we supplement features learned via a convolutional neural
network (CNN) speech encoder based on SoundStream~\cite{garbacea2019low,
zeghidour2021soundstream} with speech embeddings obtained from a pretrained Transformer model. The Transformer embeddings and encoder speech
features are concatenated after quantization---with residual vector
quantizers (VQs)~\cite{vasuki2006review}---and fed to the decoder. We
rely on GANs for the decoder design for their ability to
synthesize speech with high perceptual quality~\cite{kumar2019melgan,
kong2020hifi, zeghidour2021soundstream}. Training both the CNN encoder,
residual VQ, and the GAN-based decoder results in a noncausal speech
codec that, at a rate of $600 ~\mathrm{bps}$, reproduces speech with a
quality significantly higher than traditional or CNN-based codecs
operating at similar rates. Figure~\ref{proposed_methpd} schematically
presents our proposed codec. Some of the quality differences in
synthesized speech can be attributed to the noncausality of our codec,
and further research on causal Transformer models is required to
precisely quantify the gains resulting from speech features that exploit
long-distance dependencies.

Our work is closely related to~\cite{polyak2021speech}, which
synthesizes speech from noncausal discrete speech features, involving
speech embeddings obtained from self-supervised trained Transformer
models. In contrast to our approach, the encoder
in~\cite{polyak2021speech} includes an additional pitch
extractor~\cite{5743729} and speaker identity~\cite{heigold2016end}
models. While the authors reported high-quality speech synthesis at
$365\,\mathrm{bps}$, our choice of encoder design is more general in
that it can be adapted to any neural speech codec by augmenting the
encoder with Transformer-predicted speech embeddings. Additionally, our
proposed codec eliminates the need to have a speaker identity model, the
usage of which may lead to privacy concerns.

Our primary contribution is the reduction of the bitrate of neural
speech codecs by using speech embeddings derived from pretrained
Transformer models. In particular, we propose an end-to-end trained
speech coding approach in which the input speech is encoded by combining
Transformer embeddings with CNN encoded features. We demonstrate that
the resulting codec is capable of outperforming codecs operating at
three to four times the rate according to subjective human evaluation
metrics.

In what follows, we first describe self-supervised learning, including
the specific framework for the speech domain, and the Transformer
architecture we use. Next, we introduce our Transformer-based codec,
including details of the encoder, quantizer, and decoder modules.
Finally, we provide experimental results,
evaluating our codec with objective and subjective metrics.

\section{Self-supervised learning}\label{ssl}

With the goal of acquiring knowledge from large quantities of raw speech,
self-supervised learning involves performing a proxy task~\cite{oord2018representation} via a neural
network over the raw dataset. This task typically entails predicting the
components of the input that have been masked~\cite{baevski2020wav2vec}. In order to accomplish
this task more effectively, the neural network model must extract high-level features from contextual data
that inform the model about the masked components. Due to their
inductive bias, Transformers, which have large effective receptive fields by
design~\cite{vaswani2017attention}, are often employed as neural network models for the purpose of
self-supervised learning. Following training, the learned embeddings are used to perform various downstream
tasks~\cite{baevski2020wav2vec, zhang2020pushing, chung2021w2v}. Before
presenting the self-supervised learning framework aimed at speech, we
introduce the architecture for the Transformer model.

\subsection{Transformer architecture}\label{architecture}

We use the Transformer model introduced by~\cite{zhang2020pushing},
which consists of two sub-networks: (1) a feature encoder network that
maps input raw speech to latent speech representations; and (2) a
context network that predicts context representations given the latent
representations. The feature encoder converts the $16\,\mathrm{kHz}$ raw
input speech into a log mel-spectrogram, which is subsequently passed to
a series of subsampling convolutional layers. The overall subsampling
rate of the feature encoder network is $640$, which results in latent
speech representations with a $25\,\mathrm{Hz}$ frame rate. This design
aims to reduce the high dimensionality of speech signals, to facilitate
the learning of long-distance dependencies between the input features
more easily~\cite{baevski2020wav2vec}. The latent speech representation
is then passed to the context network, consisting of a linear layer,
followed by a stack of $24$ Conformer blocks~\cite{gulati2020conformer},
each of which is a series of multi-headed self
attention~\cite{vaswani2017attention}, depth-wise convolution, and
linear layers. With this architecture, the Transformer outputs speech
embeddings of size $1024$ with a $25\,\mathrm{Hz}$ frame rate.

\subsection{wav2vec 2.0---a self-supervised learning
framework}\label{wav2vec}

The training objective in wav2vec 2.0 is based on predicting certain
masked components in the latent speech space. To achieve this, the
masked latent speech representations are passed to the context network
to yield predicted context representations. The objective of the
Transformer is to correctly predict target context representations,
which are the result of applying a linear layer to unmasked latent
speech representations~\cite{zhang2020pushing}. To train the
Transformer, we apply a modified wav2vec 2.0 pretraining procedure that
was introduced by~\cite{zhang2020pushing}, which minimizes a contrastive
loss~\cite{oord2018representation} between the predicted and target
context representations in the masked positions. This ensures that the
context representation associated with the masked portion is accurately
predicted while being dissimilar to other target context
representations. After pretraining, the speech embeddings extracted from
the Transformer models can be used to solve downstream
tasks~\cite{polyak2021speech, baevski2020wav2vec, zhang2020pushing,
oord2018representation}, including speech coding. In the next section, we explore speech coding
as a downstream task and describe how to exploit the learned speech
embeddings for designing an low-bitrate speech codec.

\section{Transformer-based speech codec}\label{building-blocks}

Speech codecs are usually composed of three components: an encoder, a
quantizer, and a decoder. The encoder takes raw speech as an input and
extracts low-rate features that contain sufficient information to
reconstruct the speech. For a given bitrate, the quantization module
finds discrete representations of the inherently continuous encoded
features. Lastly, the decoder reconstructs the input speech signal from
the discrete encoded features. In the following sections, we describe
the encoder, quantizer, and decoder modules of the proposed speech
codec.

\subsection{Encoder}

We use speech embeddings derived from a pretrained Transformer model,
$E_T: \mathcal{X} \to \mathcal{E}_T$ for speech coding, where
$\mathcal{X}$ and $\mathcal{E}_T$ are the raw speech and Transformer
embeddings spaces, respectively. The Transformer takes raw speech,$\B{x}
\in \mathcal{X}$, as input and extracts low-rate speech features,
$E_T(\B{x}) \in \mathcal{E}_T$. We use the $21\,\mathrm{st}$ Conformer
block for embedding extraction as it results in higher quality
synthesized speech. Similar observations have been made for other
speech-related downstream tasks~\cite{pasad2021layer,
shor2021universal}. To encode other sources of information that the
Transformer embeddings potentially lack~\cite{polyak2021speech}, we
concatenate the embeddings with features learned from a CNN encoder
based on SoundStream~\cite{zeghidour2021soundstream}. This encoder,
denoted by $E_C: \mathcal{X} \to \mathcal{E}_C$, takes raw speech as
input and generates $64$ dimensional speech features with a frame rate
of $50\,\mathrm{Hz}$.

\subsection{Quantizer}\label{ResidualVQ}

Transmission of continuous speech features over low-bandwidth channels
is achieved via VQs~\cite{vasuki2006review}, where the features are
turned into discrete representations while introducing minimal
distortion. To prevent the requirement to store a very large codebook,
we utilize residual VQs~\cite{zeghidour2021soundstream,
vasuki2006review} in which the allotted codebook size is distributed
among a cascade of VQs. This approach has the advantage of increasing
the bitrate for a fixed actual codebook
size~\cite{zeghidour2021soundstream}. In this approach, each VQ uses
the quantization error of the preceding VQ as input with the first VQ
inputting the original feature vector. We use two independent residual
VQs for quantizing the Transformer embeddings and CNN features. To
improve the quantization efficiency of the Transformer embeddings, we
reduce their dimensionality via a $1024 \times 64$ linear layer. In the
rest of the paper, $Q(\cdot)$ denotes the quantizer module, involving
two residual VQs and the dimensionality reduction operator. Quantization
is inherently non-differentiable. Therefore, during minimization of loss
functions that involve quantization, we define the gradient of
$Q(\B{x})$ with respect to $\B{x}$ as identity. While this introduces
errors in gradients, empirical
observations~\cite{zeghidour2021soundstream} suggest that the error does
not prohibit the optimization procedure from providing reasonable
results.

\subsection{Decoder}\label{GANs}

Following quantization, the decoder synthesizes the original speech
signal. In this work, we adapt the GAN-based decoder proposed
in the SoundsStream coder~\cite{zeghidour2021soundstream}, motivated by its ability to
synthesize high perceptual quality speech. We replicate the
$25\,\mathrm{Hz}$ quantized Transformer embeddings to match the
$50\,\mathrm{Hz}$ frame rate of CNN encoder features. After
concatenation, these features are passed to the generator network, $G:
\mathcal{E}_T \times \mathcal{E}_C \to \mathcal{X}$, which aims to map
quantized speech features back to the speech domain. Our adversarial
training framework (cf. Section~\ref{training}) relies on two types of
discriminators, time-domain and short-time Fourier transform
(STFT) discriminators. The STFT
discriminator~\cite{zeghidour2021soundstream}, $D_0$, uses the STFT of
the speech as input with real and imaginary parts as separate channels.
On the other hand, three wave domain
discriminators~\cite{zeghidour2021soundstream, kumar2019melgan}, denoted
by $D_k,\, k=1,2,3$, use the speech signal at different resolutions,
e.g., original signal and subsampled by factors of two and
four, as their input. This allows them to learn the characteristics of
speech signal at different time scales, improving the
perceptual quality of the synthesized speech~\cite{zeghidour2021soundstream}.

\section{Training}\label{training}

We train the parameters involved in our proposed codec in an
end-to-end fashion, except for the pretrained Transformer weights, which
are kept fixed. The optimization problem consists of minimizing a series
of loss functions to determine the codec parameters
$\boldsymbol{\theta}$, of the CNN encoder, the
dimensionality reduction operator, the residual VQs, and the generator
network, as well as the weights for the discriminators,
denoted by $\boldsymbol{\phi}$. Note that the Transformer is fixed and
we do not finetune it. The resulting loss function is the weighted sum
of the loss functions described below, where the weights are
hyperparameters.

\subsection{Adversarial loss}\label{adv}

GANs are known for their ability to generate speech with high perceptual
quality~\cite{zeghidour2021soundstream, kumar2019melgan}. Their success
is partially attributed to the adversarial loss, which involves training
discriminators to guide the generator network to produce high quality
speech. In this work, we use a hinge GAN loss
function~\cite{lim2017geometric} given by,
\begin{equation}
  \mathcal{L}^{\text{adv}}_{\boldsymbol{\theta}}=\BB{E}_{\B{x}}
  \left[\frac{1}{4} \sum_{k=0}^3 \frac{1}{T_{k}} \max
  \Big(0,1-D_{k}\circ G \circ Q \circ E(\B{x})\Big)\right],
  \label{g_loss}
\end{equation}
where $E$ is the encoder, $Q$ the quantizer module, $T_k,\, k=0, \ldots
3,$ the number of logits at the $D_k$ output along the time dimension,
and $\circ$ the composition operator. The loss function (\ref{g_loss})
is used to update the parameter $\boldsymbol{\theta}$. On the other
hand, the discriminators loss function for updating $\boldsymbol{\phi}$
is
\begin{equation}
  \begin{aligned}
    \mathcal{L}^{\text{adv}}_{\boldsymbol{\phi}}=
    \BB{E}_{\B{x}} & {\left[\frac{1}{4}
    \sum_{k=0}^3 \frac{1}{T_{k}} \max \Big(0,1-D_k(\B{x})\Big)\right]}
    \\
    + \BB{E}_{\B{x}} & {\left[\frac{1}{4} \sum_{k=0}^3 \frac{1}{T_{k}}
    \max \Big(0,1+D_{k}\circ G \circ Q \circ E(\B{x})\Big)\right]. }
  \end{aligned}
  \label{d_loss}
\end{equation}

\subsection{Feature matching loss}\label{feature_match}

In addition to the output of the discriminators, we use their feature
maps, i.e., values in the intermediate layers, to construct a loss
function. This loss, known as the feature matching
loss~\cite{kumar2019melgan}, minimizes the difference in discriminator
features maps for real and synthesized input speech. This difference,
usually calculated as an $\ell_1$ distance, acts as a learned metric
function and has proven useful for generating high quality speech with
GANs~\cite{kumar2019melgan}. The loss function is given by
\begin{equation}
\begin{aligned}
  & \mathcal{L}_{\boldsymbol{\theta}}^{\text{feat}} = \\
  & \BB{E}_{\B{x}} \bigg[ 
  \sum_{k=0}^4\sum_{l=1}^L \frac{1}{4LT_{k, l}}
  \Big \| D_{k}^{(l)}(\B{x})
  - D_{k}^{(l)} \circ G\circ Q \circ E(\mathbf{x})
   \Big \|_1\bigg],
  \label{feat_loss}
\end{aligned}
\end{equation}
where $l$ denotes the layer index of the discriminators.

\subsection{Reconstruction loss}\label{rec_loss}

To ensure fidelity of the synthesized speech with respect to the encoder
features, we enforce the input and synthesized speech signals to be
coherent in the mel-spectrogram domain by
minimizing~\cite{zeghidour2021soundstream}
\begin{equation}
  \begin{aligned}
    \mathcal{L}_{\boldsymbol{\theta}}^{\text{recon}}=&
    \sum_{s \in \{ 2^i \mid i = 6, \ldots, 11 \} } \sum_{t}\Big\|
    \mathcal{S}_{t}^{s}(x)
    -\mathcal{S}_{t}^{s}\circ G \circ Q \circ E(\mathbf{x})\Big\|_{1}\\
    & + \sqrt{\frac{s}{2}} \sum_{t}\Big\|\log \mathcal{S}_{t}^{s}(\B{x})
    -\log\mathcal{S}_{t}^{s}\circ G \circ Q \circ E(\mathbf{x})\Big\|_{2},
  \end{aligned}
  \label{recon_loss}
\end{equation}
where $S_{t}^{s}$ denotes the $t^{\text{th}}$ frame of a $64$-bin
mel-spectrogram computed with window length $s$ and hop length $s/4$.

\subsection{Quantization loss}\label{q_loss}

To train the residual VQ in the context of our speech codec, we
initialize the codebooks via running k-means on a batch of feature
vectors, i.e., $E(\B{x})$ for a batch of $\B{x}$, as advocated
by~\cite{zeghidour2021soundstream}. Here $E$ represents either $E_T$ or
$E_C$. The update of codebooks during training involves: (1) bringing
the codebook closer to the output of the encoder $E(\B{x})$ via the
codebook loss and (2) encouraging the output of the encoder to create
features close to the codebook via the commitment
loss~\cite{NIPS2017_7a98af17}. Note that, as for $E_T$, the Transformer
weights are kept fixed and only the dimensionality reduction operator is
optimized. By combining these two loss functions we arrive at,
\begin{equation}
  \mathcal{L}_{\boldsymbol{\theta}}^{\text{quant}}=
  \|\mathrm{sg}[E(\mathbf{x})]-\mathbf{e}\|_{2}^{2}+\beta\|
  \mathrm{sg}[\mathbf{e}]-E(\mathbf{x})\|_{2}^{2},
  \label{quant_loss}
\end{equation}
where $\mathbf{e}$ is the codebook entry nearest to $E(\mathbf{x})$,
$\mathrm{sg}$ is the stop gradient operation, which prevents the
gradients from backpropagating during optimization, and $\beta$ is a
hyperparameter.

\section{Experiments}\label{experiments}

In the experiments presented here, we compare the quality of synthesized
speech with our proposed codec with traditional or neural speech codecs
via objective and subjective metrics. We begin with describing the
details of training our codec.

\subsection{Training configurations}\label{train_details}

To create our codec modules, we adapted the architectures of the
encoder, generator, and discriminators used
in SoundStream~\cite{zeghidour2021soundstream}. In order to evaluate the performance
of the proposed codec as a function of bitrate we trained nine codecs
identical in encoder and decoder architectures, but operating at
$600\,\mathrm{bps}$, $900\,\mathrm{bps}$, and $1800\,\mathrm{bps}$,
where at each bitrate we trained three codecs with varying bitrate
allocations: (1) using only Transformer embeddings; (2) using only CNN
features; (3) using both sources of information. To adjust the total
bitrate of the codec as well as to distribute it amongst the CNN and
the Transformer, we fixed the codebook size of each VQ to $64$ but
varied the number of cascaded VQs. For the
third set of codecs, we selected the bitrate allocation between
Transformer embeddings and CNN features that results in the best
synthesized speech quality. This amounts to an even split of bitrates
for the $600\,\mathrm{bps}$ and $1800\,\mathrm{bps}$ codecs, and a
$1/3$--$2/3$ Transformer--CNN split for the codec operating at
$900\,\mathrm{bps}$. We trained these models on a training split subset of the Mozilla Common Voice 
and LibriVox datasets (approximately $700\,\mathrm{hours}$ each), by taking up to $10$ random $1.28\,\mathrm{s}$ segments per file, which discards a significant portion of longer files.  We computed Transformer embeddings over the
entire utterance samples and created batches of $256$
$1.28\,\mathrm{s}$ utterance segments and Transformer embedding pairs for training. The
training involved minimizing the combination of the loss functions
described in Section~\ref{training} with equal weights of $1.0$, except
for the features matching and quantization losses, where they are
weighted by $100.0$ and $0.4$, respectively. We used $200\,\mathrm{k}$
iterations of Adam optimizer with learning rate $10^{-4}$. The set of
weights for balancing different loss objectives as well as the
optimizer learning rate were obtained via extensive hyperparameter
tuning. For testing, we evaluated the quality of the synthesized speech
using $4\,\mathrm{s}$ long clean English speech utterances from the VCTK
dataset~\cite{yamagishi2019cstr}. We did not test our method on noisy
utterances as the data during training was not contaminated with noise.
\begin{figure}
    \centering
    \includegraphics[width=0.95\hsize]{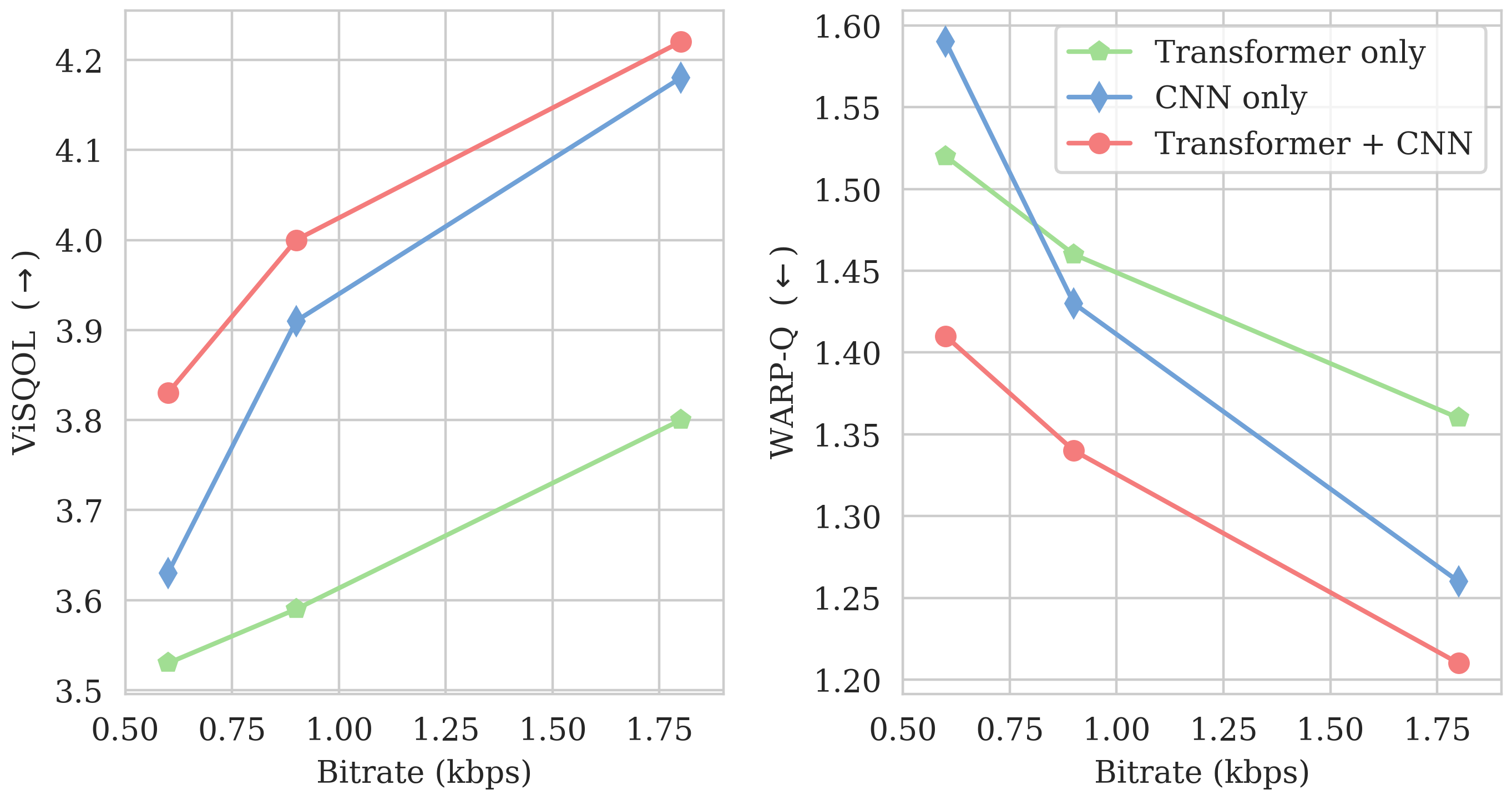}
    \caption{ViSQOL~\cite{chinen2020visqol} (left, higher is better) and
    WARP\_Q~\cite{jassim2021warp} (right, lower is better) metrics for
    three encoder designs: (1) pretrained Transformer only (green); (2)
    CNN encoder only (blue); and (3) Transformer and CNN encoder
    together (red).}
    \vspace*{-0.5cm}
    \label{objective}
\end{figure}

\subsection{Results}\label{results}

For a fixed bitrate, combining Transformer embeddings
and CNN encoder features produces higher quality speech than
the use of either Transformer embeddings or CNN encoder features alone
We computed ViSQOL~\cite{chinen2020visqol}---an
objective metric for estimating perceptual speech quality---and
WARP\_Q~\cite{jassim2021warp}---an objective metric specifically
designed for neural speech codec---shown in Figure~\ref{objective}. We
observe that our proposed codec consistently achieves higher ViSQOL and
lower WARP\_Q scores, indicating better speech quality across the three
selected bitrates. This observations suggests that the CNN encoder
learns to encode certain speech information that compliments the
Transformer embeddings in synthesizing speech. We also qualitatively
noticed that using only Transformer embeddings for speech synthesis
sometimes leads to speaker identity distortions which were absent in our proposed codec.
\begin{figure}
  \centering
  \includegraphics[width=0.75\hsize]{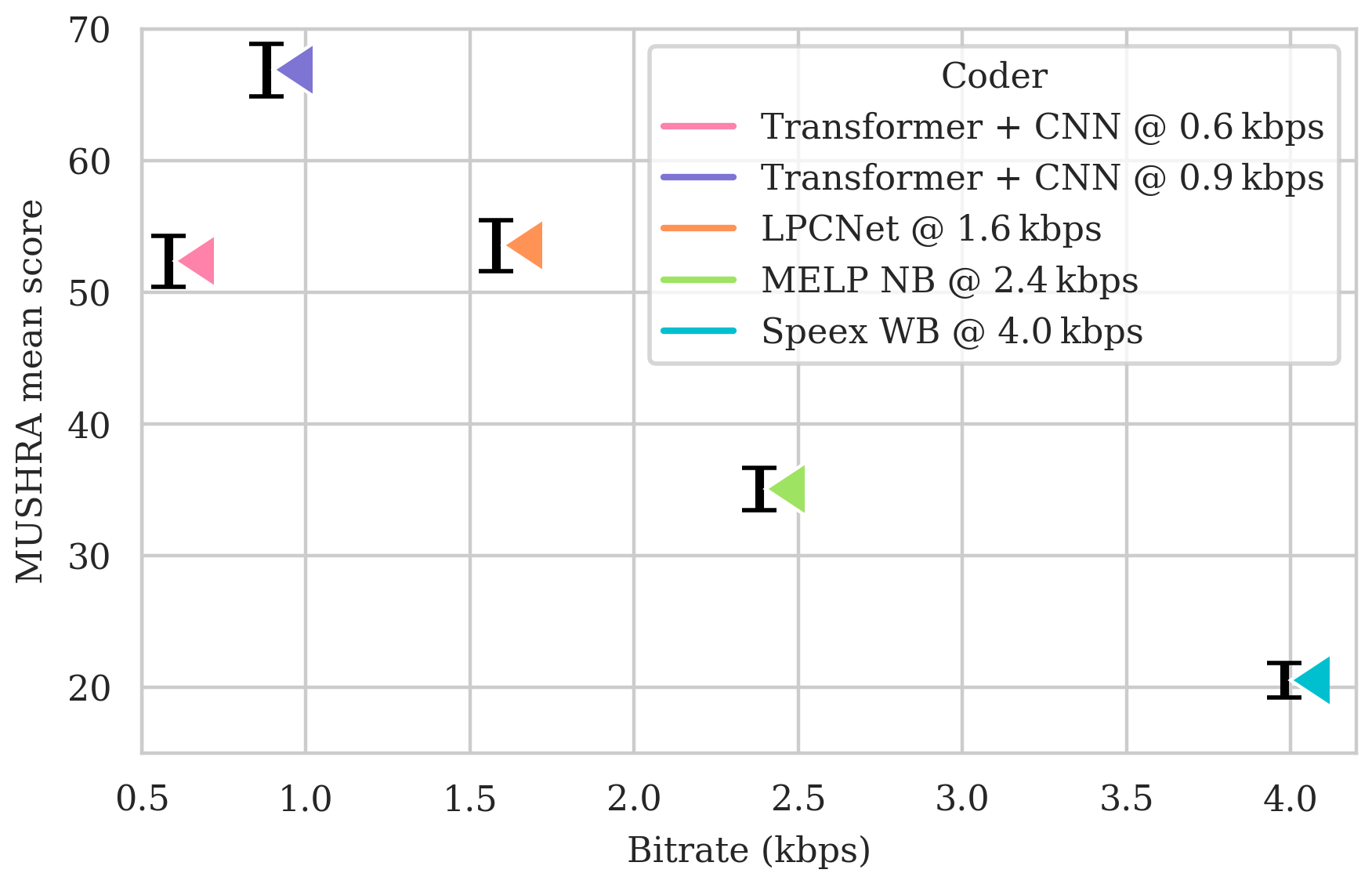}
  \caption{MUSHRA subjective test. Codecs closer to the top-left corner
  have better perceptual quality and lower bitrates. The reference with
  MUSHRA score near $100$ is not included in the figure. The indicted
  interval in black represents the $95\%$ confidence interval for each
  score.}
  \vspace*{-0.5cm}
  \label{subjective}
\end{figure}

We evaluated the perceptual quality of the synthesized speech produced
by our ultra-low bitrate codecs, operating at $600\,\mathrm{bps}$ and
$900\,\mathrm{bps}$, by comparing to Speex
wideband~\cite{valin2016speex}, MELP
narrowband~\cite{bielefeld1996developing}, and
LPCNet~\cite{valin2019lpcnet}, operating at $4\,\mathrm{kbps}$,
$2.4\,\mathrm{kbps}$, and $1.6\,\mathrm{kbps}$, respectively. We
performed MUSHRA~\cite{recommendation20011534} subjective evaluations of
our proposed codec by human raters in order to evaluate its performance.  The test used $24$ raters per utterance ($78$ raters in total) over $32$ clean speech male and female utterances.
We further post-screened the raters by removing the ones that did not rate the hidden reference above $90$ at least $80$ percent of the time.
We did not include the codecs utilizing only one of CNN features or
Transformer embeddings in this subjective test as their perceptual
quality was clearly worse than that of the proposed codec.
Figure~\ref{subjective} plots the mean MUSHRA score and $95\%$
confidence intervals as a function of codec bitrate, where higher values
indicate better quality. We observe that our $900\,\mathrm{bps}$
codec clearly outperforms all the other codecs. The $600\,\mathrm{bps}$
codec performs as well as the LPCnet codec while outperforming MELP and
Speex.

\section{Conclusions}

Reducing the bandwidth required to transmit speech while preserving the
perceptual quality remains challenging. To extract high performance from
an ultra-low-bitrate neural speech codec, we utilized the long-distance
dependencies inherent in speech signal by incorporating speech
embeddings from a pretrained Transformer in the encoding phase. Using
these speech embeddings in conjunction with speech features encoded by a
convolutional encoder yielded a noncausal speech codec capable of
operating at $600\,\mathrm{bps}$ with high perceptual quality. Our
experiments show that the proposed codec significantly outperforms the
original neural speech codec with respect to the quality of synthesized
speech when operating in the ultra-low bitrate regime. In addition, the
subjective experiments indicate comparable to or better perceptual
speech quality compared to conventional codecs operating at three to
four times the rate. Further research on causal Transformer models is
required to quantify the extent to which the gain in bitrate is related
to the incorporating long-distance dependencies in the speech signal.

\section{Acknowledgements}

We thank Marco Tagliasacchi and Neil Zeghidour for providing the SoundStream architecture and advice, and Wei Han and the speech team for providing the pretrained Transformer model.

\newpage

\bibliographystyle{IEEEtran}
\bibliography{main}

\end{document}